\documentstyle[12pt]{article}
\def\be{\begin{eqnarray}}
\def\ee{\end{eqnarray}}
\def\nn{\nonumber}
\newcommand{\ba}{\begin{eqnarray}}
\newcommand{\non}{\nonumber \\}
\newcommand{\ea}{\end{eqnarray}}
\newcommand{\re}[1]{(\ref{#1})}
\newcommand{\eq}[1]{Eq.(\ref{#1})}
\newcommand{\halpha}{\hat\alpha}
\newcommand{\ha}{\hat\alpha}
\newcommand{\hbeta}{\hat\beta}
\newcommand{\hb}{\hat\beta}
\newcommand{\bmu}{\bar\mu}
\newcommand{\hC}{\hat C}
\newcommand{\hrho}{\hat\rho}
\newcommand{\brho}{\bar\rho}
\def\la{\left\langle}
\def\ra{\right\rangle}
\newcommand{\tr}{{\rm Tr}}
\makeatletter
\def\eqnarray{\stepcounter{equation}%
              \let\@currentlabel=\theequation
              \global\@eqnswtrue
              \global\@eqcnt\z@
              \tabskip\@centering
              \let\\=\@eqncr
              $$%
 \halign to \displaywidth\bgroup
    \eqnumphantom\@eqnsel\hskip\@centering
    $\displaystyle \tabskip\z@ {##}$%
    &\global\@eqcnt\@ne \hskip 2\arraycolsep
         \hfil$\displaystyle{##}$\hfil
    &\global\@eqcnt\tw@ \hskip 2\arraycolsep
         $\displaystyle\tabskip\z@{##}$\hfil
         \tabskip\@centering
    &{##}\tabskip\z@\cr}
\def\eqnumphantom{\phantom{(\theequation)}}

\begin{document}
\noindent December, 1993
\begin{flushright}
UBC-27/93 \\
ITEP-M5/93
\end{flushright}

\bigskip

\begin{center}
{\LARGE Evaluation of Observables in the Gaussian}
\vspace{0.5cm}

{\LARGE $N=\infty$ Kazakov-Migdal Model}

\vskip 1.5truecm

{\large M.I.~Dobroliubov}$~^{a,} $\footnote{Permanent address:
Institute for Nuclear Research, Academy of Sciences of Russia, Moscow,
Russia}$^{,2}$, {\large A.~Morozov}$~^b$, {\large
G.W.~Semenoff}$~^{a,2}$, {\large N.~Weiss}$~^{a,}$\footnote{This work
was supported in part by the Natural Sciences and Engineering Research
Council of Canada}
\vskip 1truecm

{}$^a$~{\it Department of Physics, University of British Columbia\\
Vancouver, British Columbia, V6T 1Z1 Canada} \\ \mbox{} \\ {}$^b$~{\it
Institute of Theoretical and Experimental Physics} \\ {\it 117259
Moscow, Russia}
\end{center}

\vskip 0.5 truein

\centerline{ABSTRACT}

\bigskip

\noindent
We examine the properties of observables in the Kazakov-Migdal model.
We present explicit formulae for the leading asymptotics of adjoint
Wilson loops as well as some other observables for the model with a
Gaussian potential. We discuss the phase transiton in the large $N$
limit of the $d=1$ model. One of appendices is devoted to discussion
of the $N =\infty$ Itzykson-Zuber integrals for arbitrary eigenvalue
densities.

\newpage
\setcounter{footnote}{0}
\section{Introduction}
The Kazakov-Migdal model (KMM) was introduced in \cite{KM} as an
example of a gauge-invariant matrix model on a $D$-dimensional
lattice.  It consists of an $N\times N$ Hermitean matrix scalar field
$\Phi(x)$ which which lives on sites, $x$, and transforms in the
adjoint representation of $SU(N)$ and a unitary $N\times N$ matrix
gauge field (link variable) $U(xy)$ which lives on links, $\langle
xy\rangle$.  The partition function is
\equation
Z=\int d\Phi [dU]\exp\left(-N\sum_x{\rm
Tr}V[\Phi(x)]+N\sum_{<x,y>}{\rm
Tr}\Phi(x)U(xy)\Phi(y)U^{\dagger}(xy)\right)
\label{km}
\endequation
where $V[\Phi]$ is a self-interaction potential for the scalar field.
Due to the absense of a kinetic term (such as, for example, a Wilson
term) for $U$, one can integrate out the gauge variables in \re{km}
exactly.\footnote{The exact integration of the gauge fields from
(\ref{km}) uses the Itzykson-Zuber formula \cite{itzub} and can in
some sense be regarded as a way to avoid solving the long-standing
problem of summation of planar diagrams in lattice gauge theory.} The
resulting effective scalar theory contains $ N$ degrees of freedom,
the eigenvalues of the matrix $\Phi(x)$, whose quantum fluctuations
are suppressed in the large $N$ limit.  This makes it possible to
employ large $N$ techniques. The field $\Phi(x)$ plays the role of a
master field and the density of its eigenvalues, $\rho(\phi)$, can in
principle be found by solving a saddle-point (master field) equation.
In this sense the model is regarded as being solvable in the large $N$
limit.  However, explicit solution of the master field equation
remains one of the major technical obstacles in this program.

With the assumption of a translation invariant master field, Migdal
\cite{Migdal} has reduced this problem to finding the solution of a
singular highly non-linear integral equation. Some further progress
has also been made using the method of loop equations
\cite{DMS}. In both approaches, technical problems prevent one from
obtaining very many analytic results. For instance, explicit solutions
for the spectral density of the scalar field are thus far known for
only two kinds of potentials: the Gaussian one ($V[\Phi] =
\frac{m^2}{2}\Phi^2$ ) solved by Gross \cite{Gross} and a potential
related to the Penner model, which was investigated by Makeenko
\cite{Penner}.  Furthermore, even for a Gaussian potential where the
spectral density is known explicitly, only expectation values of
products of scalar fields on a site \cite{Gross} and one-link
correlators of the gauge field \cite{DMS} have been computed so far.
In this Paper, we shall use these previous results to present some
explicit formulae for observables in the Gaussian KMM.  In particular,
we give a caluclation of the expectation value of the adjoint Wilson
loop as well as the gauge invariant two-point correlator of the scalar
field.  We regard this as an intermediate step in understanding the
continuum limit of the KMM in the general case.

It was originally hoped that the Kazakov-Migdal model would possess a
continuum limit which would be a gauge theory, such as QCD.  It was
also clear from the outset that this could not be realized in a
straightforward way.  Because the action of the model \re{km} does not
contain the kinetic term for the gauge field (this was exactly the
fact which allowed one to integrate out the gauge fields), the KMM
posesses an additional local (gauge) $Z_N$ invariance \cite{ZN} which
implies that all of the usual Wilson loops vanish unless they have
vanishing area.  This can be interpreted as an infinite string tension
and results in ``ultraconfinement'', where no propagation of color is
allowed at any distance scales.  An ordinary continuum gauge theory
does not have such a property.

Thus, if the KMM is to describe continuum gauge theories, then either
there must exist a more sophistocated continuum limit \cite{KMSW}, or
the model itself should be modified in a way which would break the
local $Z_N$ symmetry explicitly (yet preserving its solvability)
\cite{Migdal-mixed,DKSW}.

Irrespectively of its relevance for the description of $D>2$
Yang-Mills theories, the KMM has proven to be very interesting from a
purely theoretical point of view as an intermediate step between
generic matrix models which involve full unitary matrix integrals, and
the well studied ``$c<1$'' matrix models in which the unitary matrices
(``angular variables'') completely decouple \cite{morev}.  The KMM is
furthermore directly relevant to a certain sector of $c=1$ string
theory \cite{boulkaz} where it indeed has a phase transition and one
can discuss its continuum limit.

The choice of physical observables in the KMM is not entirely obvious.
Due to the existence of the additional $Z_N$ gauge invariance, all of
the conventional Wilson loops vanish. Note, that in the large $N$ limit and in
the mean field approximation, the unitary integrals in the KMM can be
reduced to single link integrals of the type
\be
{\cal C}(n|\Phi,\bar\Phi) \propto
\int_{N\times N} [dU] U_{i_1j_1}\ldots U_{i_nj_n}U^\dagger_{\hat j_1\hat i_1}
\ldots
U^\dagger_{\hat j_n\hat i_n} e^{{\rm Tr} N\Phi U\bar\Phi U^\dagger},
\label{1li}
\ee
They must involve the same number of $U$'s and $U^\dagger$'s in order
to be invariant under the local $Z_N$ transformation. The simplest
gauge invariant observables which can be constructed out of the $C$'s
are the adjoint Wilson loop and the filled Wilson loop.  Both of these
involve only the averages of $\vert U_{ij}\vert^2$ i.e. ${\cal
C}(1|\Phi,\bar\Phi)$. (As we mentioned above, even for a soluble model
such as the Gaussian KMM ${\cal C}(n|\Phi,\bar\Phi)$ have been
evaluated explicitly only for $n=0$ and $n=1$).

Before presenting these formulae we make some general comments about
the integral (\ref{1li}) for arbitrary $n$.  These integrals fall into
the general framework of applicability of the Duistermaat-Heckmann
theorem, and,in principle, they can be explicitly evaluated
(see \cite{KMSW} and references therein for details).  However,
existing explicit expressions for $n\geq 1$
\cite{MoSha} are too complicated to be used effectively when
$N\rightarrow\infty$, which is the limit of interest in the KMM. What
is required is an expression for ${\cal C}(n|\Phi,\bar\Phi)$ in terms
of the densities $\rho(\phi) \equiv
\frac{1}{N}\sum_{i=1}^N \delta(\phi-\phi_i)$ and $\bar\rho(\phi)$ of
eigenvalues of the matrices $\Phi$ and $\bar\Phi$.  This would allow
taking a smooth large $N$ limit.  In this limit, discrete sums over the
indices become continuous integrals over the eigenvalues,
$\frac{1}{N}\sum_i
\rightarrow \int d\phi\rho(\phi)$.  Expressing all quantities in terms
of densities implies also relabelling of tensor structures like ${\cal
C}(n)$ for $n\geq 1$. Indices $i,j,\ldots$ should be substituted by
$\phi_i,\phi_j\ldots$ according to the rule $
C_{\phi_{i_1}\ldots\phi_{i_m}} \equiv N^{m/2}C_{i_1\ldots i_m}$, so
that for example
\be
\sum_{j=1}^N C_{ij}C_{jk} = \frac{1}{N}\sum_{\phi_j}  C_{\phi_i\phi_j}
C_{\phi_j\phi_k} = \int d\phi \rho(\phi) C_{\phi_i\phi}
C_{\phi\phi_k}~,
\label{Cnorm}
\ee
At the moment there exist only a few methods for the evaluation of
$C(n\vert\rho,\bar\rho)$ with $n=0$ \cite{Ma} and $n=1$
\cite{DMS} (for details see Appendix B).
Though they have the potential to work in a general case, these
methods have not proven to be very efficient in practice.  Until now
the quantities $C(n \vert
\rho,\bar\rho)$ have been computed only for two cases: for $n=0,1$ for the
semi-circle \cite{Gross,DMS} and $n=1$ for the ``Penner-like''
\cite{Penner} distributions of the eigenvalues.
(The latter is more complicated and will not be considered here.)

With the assumption that the mean field is spatially homogenous, the
scalar field in the Gaussian KMM has the semi-circle spectral density
\be
 \rho(\phi)d\phi = \bar\rho(\phi)d\phi = \rho_{\mu}(\phi)d\phi
\equiv \frac{1}{\pi}
\sqrt{\mu - \frac{\mu^2\phi^2}{4}}d\phi ~~(\vert\phi\vert
\leq\frac{2}{\sqrt\mu})
\ee
For this spectral density the explicit results for the Itzykson-Zuber
integral ${\cal C}(0|\rho_\mu,\rho_\mu)$ \cite{Gross} and correlators
${\cal C}(1|\rho_\mu,\rho_\mu)$ \cite{DMS} are as follows
\ba
& C(\mu) = \lim_{\stackrel{N\rightarrow\infty}{\rho = \rho_\mu}}\frac{1}{N^2}
\log \int_{N\times N} [dU]  e^{{\rm Tr} N\Phi U\bar\Phi U^\dagger}~~~,
\nn \\
& C(\mu) = \frac{\sqrt{\mu^2+4} - \mu}{2\mu} -
\frac{1}{2}\log \frac{\sqrt{\mu^2+4} + \mu}{2\mu} ~~~;
\label{parfun} \\
& \delta_{i\hat
 i}\delta_{j\hat j} C_{ij}(\mu) =
\frac{\int_{N\times N} [dU] U_{\hat i\hat j}U^\dagger_{j i}
 e^{{\rm Tr} N\Phi U\bar\Phi U^\dagger}}{\int_{N\times N} [dU]
 e^{{\rm Tr} N\Phi U\bar\Phi U^\dagger}} ~~~, \nn \\
&  C_{\alpha\beta}(\mu) =
\lim_{\stackrel{N\rightarrow\infty}{\rho = \rho_\mu}}
N C_{ij}(\mu) =
{1\over 2} \frac{\mu+\sqrt{\mu^2+4}}
{\alpha^2 + \beta^2 - \sqrt{\mu^2+4}\alpha\beta + \mu} ~~~.
\label{Cab}
\ee
In practical calculations it is more convenient to use the
``normalized'' semi-circle distribution,
$\hat\rho_s(\hat\phi),~\hat\phi = \frac{2}{\sqrt\mu}\phi$, which has
support $[-1,1]$ and
\be
\rho_s(\hat\phi) \equiv \frac{1}{\pi}\sqrt{1-\hat\phi^2}~,~~
\rho(\phi)d\phi \equiv 2\hat\rho_s(\hat\phi)d\hat\phi,
\ee
so that
\ba
&C(\tau) = \frac{\cosh\tau-\sinh\tau}{2\sinh\tau} - {1\over 2} \log
\frac{\cosh\tau+\sinh\tau}{2\sinh\tau}~~~,
\non
& \hat C_{\hat\alpha\hat\beta}(\mu) =
2C_{\alpha={\sqrt{\mu}\over2}\hat\alpha\, , \beta={\sqrt{\mu}\over
2}\hat\beta} = \frac{\sinh\tau(\cosh\tau + \sinh\tau)} {\hat\alpha^2 +
\hat\beta^2 - 2\hat\alpha\hat\beta\cosh\tau + \sinh^2\tau}~~~,
\label{Cabnorm} \\
& {\rm where} \sinh\tau \equiv \frac{\mu}{2},\ \ \ \cosh\tau =
\frac{\sqrt{\mu^2
+4}}{2} ~~.
\nn
\ea
(sometimes we shall use notation $\hat C_{\hat\alpha\hat\beta}(\tau)$
instead of $\hat C_{\hat\alpha\hat\beta}(\mu)$, which should not cause
any confusion).

\section{Observables in the KMM}

The set of observables in the KMM is restricted by the local
$Z_N$-invariance, which, for example, forces averages of all the
fundamental-representation Wilson loops to vanish and results in
ultraconfiment.  Non-vanishing observables contain an equal number of
$U$ and $U^\dagger$ matrices at every link,\footnote{ Since we
consider the large-$N$ limit, we neglect ``baryonic'' observables,
containing $N$ or more $U$-matrices per link without any
$U^\dagger$'s} examples are shown in Fig.1. Dots can appear at every
site and denote insertion of any number of $\Phi$ or $\bar\Phi$
operators. One can account for all possible insertions by using the
generating function $\frac{1}{\lambda_xI -\Phi}$ at every site.
Occasionally some double-lines can become quadruple etc at some links,
then the knowledge of higher-order correlators ${\cal C}(n)$ with
$n\geq 2$ is required for evaluation of averages. We restrict
ourselves to consideration of the non-degenerate situations, where
double lines never overlap. In such cases the answer is further
simplified
\cite{KMSW} by the  occurence of $\delta_{i\hat i}\delta_{j\hat
j}$ in eq.(\ref{Cab}). It allows us to consider every observable as
associated with a graph $\Gamma$ which ascribe a single index $i_x$ to
every site (i.e. effectively substituting double lines by single lines,
with $C_{ij}$ playing the role of the propagator, and $(\lambda_x -
\phi_{i_x})^{-1}$ - that of the vertex):
\be
\langle {\cal O}_{\Gamma} \rangle = N^{-n_\Gamma}
\sum_{\{i_x, x\in\Gamma\}} \left(\prod_{x\in\Gamma}
\frac{1}{\lambda_x - \phi_{i_x}} \prod_{\langle xy \rangle \in\Gamma}
C_{i_xi_y}\right) \ \stackrel{N\rightarrow\infty}{\longrightarrow} \nn \\
 \rightarrow N^{-n_\Gamma + \#(x\in\Gamma) - \#(\langle xy \rangle \in\Gamma)}
\prod_{x\in\Gamma} \int\frac{\rho(\phi_x)d\phi_x}
{\lambda_x - \phi_{i_x}}\prod_{\langle xy \rangle \in\Gamma}
C_{\phi_x \phi_y}~,
\label{Nfac}
\ee
where, as usual, $n_\Gamma$ is the number of traces, contained in the
definition of ${\cal O}_{\Gamma}$ (for discussion of different
normalization prescriptions see ref.\cite{KMSW}).  Note that, with our
normalization of $C$ and $\rho$, each site contributes a factor of $N$
and each link a factor of $1/N$.  It will turn out that the product of
integrals in the right-hand-side of (9) is well-defined and
$N$-independent.  With this normalization of operators, the only ones
which are non-zero in the infinite $N$ limit are tree-like
configurations - these always have one index sum ($n_{\Gamma}=1$) and
also $\#(x\in\Gamma) - \#(\langle xy \rangle \in\Gamma)=1$. Every loop in a
configuration will only increase $\#(x\in\Gamma) - \#(\langle xy \rangle
\in\Gamma)$, making it greater than one.

At least in principle, the integral on the r.h.s.  can be evaluated
for any graph, provided $\hat C_{\phi_x \phi_y}$ is known for the
given $\rho$. At the moment, however, an explicit expression is
available only for $\rho$ given by the semi-circle distribution.  In
$D=1$ there are only two types of allowed graphs: segments and circles
of length $L$ which, as we shall show, can be relatively easily
computed for the semi-circle distribution\footnote{ For lattices which
form ``rare nets'', so that the theory remains essentially
1-dimensional, more sophisticated graphs are also of interest. We,
however, ignore this possibility.}.  This is the subject of the next
section. In higher dimensions the full set of observables contains
more objects, like filled Wilson loops \cite{KMSW}, which are not
considered in the present paper.

\section{Evaluation of Observables}

\subsection{The Adjoint Wilson Loop}

The Wilson Loop is obtained by considering the trace of the
path-ordered product of the link operators for the links which occupy
a contour $\Gamma$ on the lattice
\begin{equation}
W[\Gamma]\equiv {\rm\bf tr}\prod_{<xy>\in\Gamma} U(xy)
\end{equation}
Because of the $Z_N$ symmetry, the expectation value of this operator
vanishes unless $\Gamma$ has vanishing area.  An example of an
operator which is $Z_N$ invariant and also a physical interpretation
is the adjoint Wilson loop,
\begin{equation}
W_A[\Gamma]= \frac{1}{N^2}\left(\vert W[\Gamma]\vert^2-1\right)
\end{equation}
The appearance of the constant term in this definition can be
understood if one recalls\footnote{We thank Yuri Makeenko for a
discussion of this point.} that the product of two fundamental
representation link operators contains both the adjoint and scalar
representations, the latter of which must be subtracted:
$(U_{ij}U^{\dagger}_{kl})_A=
U_{ij}U^{\dagger}_{kl}-\delta_{il}\delta_{jk}$.

In the mean field approximation of the KMM, the adjoint Wilson loop
can be evaluated using the one-link expectation values of the the
products $C_{ij}=\langle UU^{\dagger}\rangle$,
\begin{equation}
\langle W_A[\Gamma]\rangle= \frac{1}{N^2} \left({\rm\bf
tr}\left(C^{L[\Gamma]}\right)-1\right)=\frac{1}{N^2}\left(\sum_i
c_i^{L[\Gamma]}-1\right)
\label{eig}
\end{equation}
where $c_i$ are the eigenvalues of the matrix $C_{ij}$ and $L[\Gamma]$
is the length of the contour $\Gamma$.

{}From their definition it is easy to see that the matrices $C_{ij}$ are
real, have positive entries and, when the master field is homogeneous,
they are also symmetric.  Furthermore, they obey the sum rule
\begin{equation}
\sum_{i=1}^N C_{ij} =1
\label{sumrule}
\end{equation}
This sum rule implies that, for any $N$, $C_{ij}$ always has one
eigenvalue which is one, with eigenvector $(1,1,1,\ldots,1)$.

It is also easy to see that all other eigenvalues lie in the interval
$[-1,1]$. Indeed, let us consider the set of real matrices with
positive entries, satisfying the sum rule \re{sumrule}.  Obviously,
all matrices from this set have their traces bounded from above by
$N$. Further, this set is closed under the matrix multiplication: if
matrices $A$ and $B$ belong to this set, then so does the matrix $AB$.
Take now any matrix $C$ from this set and consider its power $C^n$
with $n$ even. If there are eigenvalues of the matrix $C$, absolute
value of which is greater than 1, then for big enough $n$ one can make
the trace of $C$ arbitrary large, thus violating the mentioned upper
bound. Q.E.D.

Let us recall that a rectangular adjoint Wilson loopp (whose lengthes in the
space and time directions are $L$ and $T$, respectively) in the limit
$T\rightarrow\infty$
 can be interpreted as the energy of a pair of mesons with
separation $L$,
\begin{equation}
E[L]=\lim_{T\rightarrow\infty} \frac{1}{T}\ln
\left(N^2<W_A[\Gamma]>\right)
\end{equation}
Once 1 is subtracted in the sum on the right-hand-side of (\ref{eig}),
the energy is dominated by the largest remaining eigenvalue of $C$.

In the following we shall compute $W_A$ in the large $N$ limit for the
Gaussian Kazakov-Migdal model.  First, however, let us consider the
case of $N=2$ which was previously analyzed in \cite{KMSW}.  There,
even though the mean field approximation for the scalar field is
uncontrolled (since $N=2$ is not large), it has been argued
\cite{brook} that mean field theory can give accurate results.  If the
mean field eigenvalues of the $2\times2$ scalar matrix field $\phi$
(which can be taken as traceless) in this case are $\bar\phi$ and
$-\bar\phi$, it was shown in \cite{KMSW} that
\begin{eqnarray}
C_{11}=C_{22}= \frac{1-1/4\bar\phi^2+e^{-4\bar\phi^2}
/4\bar\phi^2}{1-e^{-4\bar\phi^2}}
\nonumber\\
C_{12}=C_{21}=\frac{1/4\bar\phi^2-e^{-4\bar\phi^2}
(1+1/4\bar\phi^2)}{1-e^{-2\bar\phi^2}}
\end{eqnarray}
The eigenvalues are 1 and $\coth(2\bar\phi^2)-1/2\bar\phi^2$.
Note that the second eigenvalue varies between zero and one as
$\bar\phi^2$ goes from zero to infinity.  The energy of the meson pair
is
\begin{equation}
E=-2\ln\left(\coth(2\bar\phi^2)-1/2\bar\phi^2\right)
\label{ener}
\end{equation}
which is positive and goes to zero at the ``critical'' value of
$\bar\phi^2\rightarrow\infty$.  Note that, in this approximation the
mesons have no interaction energy, i.e. (\ref{ener}) can be
interpreted as twice the meson mass.

Now let us proceed to the large $N$ limit in the case of a Gaussian
potential. We start with the basic relation, proven for the
semi-circle distribution in Appendix A:
\be
\int_{-1}^{+1}
\hat C_{\hat\alpha\hat\beta}(\tau_1)\hat C_{\hat\beta\hat\gamma}(\tau_2)
\hat\rho_s(\hat\beta)d\hat\beta =
\hat C_{\hat\alpha\hat\gamma}(\tau_1 + \tau_2).
\label{CC-C}
\ee
Therefore for the adjoint Wilson loop of length $L$ we find
\ba
& \langle W_A[\Gamma]\rangle = {1\over N^2}\left( \int d\alpha_1
\rho(\alpha_1) \ldots d\alpha_{L}
\rho(\alpha_{L}) C_{\alpha_1\alpha_2}(\tau) \ldots
C_{\alpha_{L-1}\alpha_{L}}(\tau) -1\right)=
\non
&= {1\over N^2}\left( \int d\halpha_1 \hrho(\halpha_1) \ldots d\halpha_L
\hrho(\halpha_L) \hC_{\halpha_1\halpha_2}(\tau) \ldots
\hC_{\halpha_{L-1}\halpha_L}(\tau)-1\right) =
\non
&= {1\over N^2}\left( \int d\halpha \hrho(\halpha)
\hC_{\halpha\halpha}(L\tau)-1\right)
\non
&= {1\over N^2} \frac{e^{-L\tau}}{1-e^{-L\tau}}
\label{001}
\ea
Note that, as noted in the previous section, the normalization of the
adjoint Wilson loop is such that it vanishes in the large $N$ limit.
We can regard (\ref{001}) as the leading asymptotics.

The interaction potential for a pair of mesons is obtained from the
free energy,
\begin{equation}
E(L)=2\tau
\label{yuk}
\end{equation}
which one can interpret as just the sum of the meson masses (each equal to
$\tau$) without any interaction between mesons.

There is an apparent phase transition at $\tau=0$.
Recall that \cite{Gross}, for the Gaussian potential
$V[\phi]=\frac{m^2}{2}\phi^2$, so that the ``bare mass'' of $\phi$ is
$m^2-2D$, and $\tau$ is given by
\begin{equation}
\sinh\tau=\frac{m^2(D-1)+ D\sqrt{m^4-4(2D-1)}}{2(2D-1)}
\end{equation}
or, equivalently,
\begin{equation}
m^2=2De^{-\tau}+2\sinh\tau\approx 2D+2(1-D)\tau~~(\tau\sim0)
\end{equation}
As has been observed by Gross \cite{Gross}, the phase transition can
be approached through the physical region, $m^2\geq 2D$, only when
$D\leq 1$.

\subsection{Other Observables}

It is straightforward to compute the expectation values of a variety
of other observables.  Here, we shall present a few examples.

It is very easy to find the expectation value of powers of the scalar fiels
$\Phi$ on a single site, using the generating function
\be
E_\lambda \equiv \langle {\rm \bf tr} \frac{1}{\lambda I -\Phi}\rangle =
\sum_{n=0} \frac{\langle {\rm \bf tr}\Phi^n\rangle}{\mu^{n+1}}~~.
\ee
This generating function for the semi-circle spectral density was found by
Gross \cite{Gross} (in this case odd momenta are
vanishing since the spectral density is even)
\be
E_\lambda\equiv\int
\frac{\rho_s(\beta)d\beta}{\lambda-\beta} = \frac{\mu\lambda}{2} -
\sqrt{\frac{\mu^2\lambda^2}{4}-\mu} = \frac{1}{\lambda} \left( 1 +
\sum_{k\geq 1}\frac{(2k-1)!!}{(k+1)!}
\left(\frac{2}{\mu\lambda^2}\right)^k\right)~~,
\label{EE}
\ee
so that $\langle {\rm \bf tr}\Phi^2\rangle=1/\mu$ and so on.

 For an
adjoint loop of the length L with the insertion of the operator
$\frac{1}{\lambda I-\phi}$ into one of the products $\prod U$ at the
single site we obtain:
\ba
&\langle {\cal O}(\circ L)\rangle =  {1\over N^2}
\int d\alpha_1 \rho(\alpha_1) \ldots d\alpha_L \rho(\alpha_L)
\frac{1}{\lambda-\alpha_1}
 C_{\alpha_1\alpha_2}(\tau) \ldots C_{\alpha_{L}\alpha_1}(\tau)=
\non
&= {1\over N^2}\frac{\sqrt\mu}{2}\int d\halpha
\hrho(\halpha)\frac{1}{\hat\lambda-\hat\alpha} \hC_{\halpha\halpha}(L\tau)
\non
&= {1\over N^2} \frac{e^{-L\tau}}{1-e^{-L\tau}}
\frac{1}{\lambda^2-\frac{4}{\mu}\cosh^2(L\tau/2)}
(\lambda-\frac{1}{\mu}(e^{L\tau}+1)E_\lambda)
\label{01}
\ea
where $E_\lambda$ is defined in the Appendix.  The coefficient in
front of $\lambda^{-1}$ in the large-$\lambda$ expansion of (\ref{01})
reproduces (\ref{001}).

Similarly, for a segment with $\phi$-insertions at its ends:
\ba
&\int d\alpha_1 \rho(\alpha_1) \ldots d\alpha_L \rho(\alpha_L)
\frac{1}{\lambda-\alpha_1}\frac{1}{\nu-\alpha_L}
C_{\alpha_1\alpha_2}(\tau) \ldots
C_{\alpha_{L-1}\alpha_L}(\tau)=
\non
&= \frac{\mu}{4}\int d\halpha
\frac{\hrho(\halpha)}{\hat\lambda-\alpha}
d\hb\frac{\hrho(\hb)}{\hat\nu-\hb} \hC_{\halpha\hbeta}(L\tau)
\non
&=\frac{\lambda E_\nu+\nu E_\lambda+2\sinh (L\tau) +(\cosh(L\tau)+\sinh(L\tau))
(\frac{2\sinh(L\tau)}{\mu}E_\nu E_\lambda-\lambda E_\lambda -\nu E_\nu
)}{\nu^2+\lambda^2-2\cosh(L\tau)\nu\lambda+\frac{4}{\mu}\sinh^2(L\tau)}
\non
\label{02}
\ea

It is worth noting that, for real $\tau$, the expressions \re{01}
and \re{02} are never singular for any $\lambda$ and $\nu$, as they
should.

Also, note that this quantity is of order one, rather than $1/N^2$.
This is a result of the fact that this observable is tree-like,
whereas the previous two, which we considered, had loops (cf. the discussion
after the equation \re{Nfac}).  The gauge
invariant $\phi-\phi$ correlator, which is a flux tube with the field $\Phi$ at
the edges, can be extracted from this result by
taking the leading, order of $1/\hat\lambda^2$, $1/\hat\nu^2$, asymptotics
of (\ref{02}) as
\begin{equation}
\langle {\rm \bf tr} \; \phi\prod_{\Gamma} U \,
\phi \, \prod_{-\Gamma}U^\dagger \; \rangle = {1\over \mu}e^{-L\tau}
\label{tube}
\end{equation}
We see, that it has very simple form, with $\tau$ being the
correlation length and agrees qualitatively with our interpretation of
the result (\ref{yuk}). When the length $L$ equals zero, so that the flux tube
vanishes shrinking to a point, the expression \re{tube} reproduces the result
for the average of ${\rm \bf tr} \Phi^2$.

\section{Conclusion}

The effective field theory for the eigenvalues of the scalar field
$\Phi$ is classical in the large $N$ limit, and the classical
configurations are obtained by solving the saddle point equation.  It
is a remarkable feature of the Gaussian Kazakov-Migdal model that, in
addition to determining the classical scalar field, one can take the
fluctuations of the gauge fields into account exactly.

{}From the trivial counting of powers of $N$ we find that, with their
conventional normalization, the only operators with non-zero
expectation value in the limit $N\rightarrow\infty$ are those of
tree-like configurations of tubes of glue with powers of the scalar
fields inserted at the ends of the branches.

While the expectation value of the Wilson loops with non-zero area is
always zero in this model, the expectation value of the adjoint Wilson
loop is proportional to $1/N^2$ and obeys the perimeter law, with the
correlation length equal to $1/\tau$, which is related to the bare
scalar mass $m^2-2D$ and the coordination number of the lattice $2D$
according to (21),(22). In the continuum limit (which is possible only
in $D=1$), this corresponds to non-interacting mesons with the mass
equal to $\tau$.

For a flux tube with scalar field at each end, the expectation value
is proportional to the exponential of the length of the tube divided
by the correlation length, which is again given by $1/\tau$.  We also
obtained the general formula for the expectation value of the flux
tube with arbitrary powers of the scalar fields at each end.

\vskip 1cm

\centerline{\bf APPENDIX A}

\bigskip

This appendix contains some formulas, relevant to the proof of
eq.(\ref{CC-C}) as well as for other calculations with semi-circle
distributions.

We repeat the definition of the ``normalized'' generating functional
for the averages $\langle {\rm Tr}\Phi^{2k}\rangle$
\ba
& \hat E_{\hat\gamma}\equiv\int
\frac{\hat\rho_s(\hat\beta)d\hat\beta}{\hat\gamma - \hat\beta} =
\hat\gamma - \sqrt{\hat\gamma^2 - 1} = \frac{1}{2\hat\gamma}
\left( 1 + \sum_{k\geq 1}\frac{(2k-1)!!}{(k+1)!}
\left(\frac{1}{2\hat\gamma^2}\right)^k\right)~,
\non
& \hat E_{\hat\gamma} = \cosh\theta - \sinh\theta~, ~~~~~{\rm
where}~\hat\gamma\equiv\cosh\theta~~.
\nn
\label{E}
\ee

Let us rewrite the correlators \re{Cabnorm} for the ``normalized''
spectral density in the form
\be
\hat C_{\hat\alpha\hat\beta} = \frac{\sinh\tau(\cosh\tau + \sinh\tau)}
{\hat\alpha^2 + \hat\beta^2 - 2\hat\alpha\hat\beta\cosh\tau +
\sinh^2\tau} =
\frac{\sinh\tau(\cosh\tau + \sinh\tau)}
{(\hat\beta - \alpha_+)(\hat\beta - \alpha_-)},
\ee
where\footnote{Note that since $\ha^2\le 1$, $t$ is complex}
\be
\alpha_\pm  = \hat\alpha\cosh\tau \pm \sqrt{\hat\alpha^2 -1}
\sinh\tau = \cosh(\tau \pm t) ~~,~~~~\hat\alpha \equiv \cosh t
\label{root}
\ee
{}From \re{E} and \re{root} we get
\be
E_{\alpha_\pm} = \cosh(\tau \pm t)-\sinh(\tau \pm t)
\ee
(here the choice of the signs corresponds to that of \eq{EE})

The first applicaiton of this convenient parametrization is to check
explicitly the normalization condition \re{Cnorm} (we denote
$\alpha\equiv\cosh \theta$)
\ba
&\int d\beta \rho(\beta) C_{\alpha\beta} = 2\int d\hbeta
\hrho(\hbeta) C_{\halpha\hbeta} = \sinh\tau
(\cosh\tau+\sinh\tau)
\frac{E_{\alpha_+}-E_{\alpha_-}}{\alpha_--\alpha_+} =
\non
& =\sinh\tau (\cosh\tau+\sinh\tau) \frac{
\cosh(\tau+\theta)-\sinh(\tau+\theta)
-\cosh(\tau-\theta)+\sinh(\tau-\theta)}{\cosh(\tau-\theta)-\cosh(\tau+\theta)}
= 1
\nn
\ea
Let us now turn to the convolution of two $C$'s.
\ba
&\int d\hb \hrho(\hb) \hC_{\hb_1\hb}(\tau_i)\hC_{\hb\hb_2}(\tau_j) =
\frac{\sinh\tau_i (\cosh\tau_i-\sinh\tau_i)}{\beta_{+i}-\beta_{-i}}
\frac{\sinh\tau_j (\cosh\tau_j-\sinh\tau_j)}{\beta_{+j}-\beta_{-j}}
\non
& \times
\left\{
\frac{E_{\beta_{+j}}-E_{\beta_{+i}}}{\beta_{+i}-\beta_{+j}}
+\frac{E_{\beta_{-j}}-E_{\beta_{-i}}}{\beta_{-i}-\beta_{-j}} -
\frac{E_{\beta_{+j}}-E_{\beta_{-i}}}{\beta_{-i}-\beta_{+j}} -
\frac{E_{\beta_{-j}}-E_{\beta_{+i}}}{\beta_{+i}-\beta_{-j}} \right\} =
\non
& =\frac{\sinh\tau_i (\cosh\tau_i-\sinh\tau_i)}{\beta_{+i}-\beta_{-i}}
\frac{\sinh\tau_j (\cosh\tau_j-\sinh\tau_j)}{\beta_{+j}-\beta_{-j}}
\left\{ \coth\frac{\tau_i+\tau_j+\theta_1+\theta_2}{2} + \right.
\non
& \left. +\coth\frac{\tau_i+\tau_j-\theta_1-\theta_2}{2}
-\coth\frac{\tau_i+\tau_j-\theta_1+\theta_2}{2} -
\coth\frac{\tau_i+\tau_j+\theta_1-\theta_2}{2} \right\} =
\non
& =\frac{\sinh(\tau_1+\tau_2)(\cosh(\tau_1+\tau_2) +
\sinh(\tau_1+\tau_2)} {\hat\beta_1^2 + \hat\beta_2^2 -
2\hat\beta_1\hat\beta_2\cosh(\tau_1+\tau_2) + \sinh^2(\tau_1+\tau_2)}
{}~,
\nn
\ea
which proves \eq{CC-C}.

\vspace{1cm}
\centerline{\bf APPENDIX B}
\centerline{\bf DMS-Matytsin theory of the $N=\infty$ Itzykson-Zuber integrals}

\bigskip

This appendix contains description of the currently available indirect
methods to evaluate the (logarithm of the) Itzykson-Zuber integral
$\hat C(0\vert \rho,\bar\rho)$ and the first Itzykson-Zuber correlator
$\hat C_{\alpha\beta}(1\vert \rho,\bar\rho)$ for given eigenvalue
densities $\rho(\phi)d\phi$ and
$\bar\rho(\bar\phi)d\bar\phi$.\footnote{ In order to avoid confusion
we emphasize that ``bar'' does not mean complex conjugation,
$\rho(\phi)$ and $\bar\rho(\bar\phi)$ are just independent functions.
Complex conjugation will be denoted by ``*'' in what follows.}

As it often happens in the field theory, evaluation of the correlator
$\hat C_{\alpha\beta}$ \cite{DMS} is somewhat simpler than that of the
``free energy'' $\hat C$ \cite{Ma}, and the methods, used in the two
cases, though similar in many respects, are still complementary rather
than identical.

\bigskip

\centerline{{\it Evaluation of $\hat C_{\alpha\beta}$} \cite{DMS}}

\bigskip

Despite it produces the answers for individual (one-link)
Itzykson-Zuber correlators the method of \cite{DMS} actually works
within the framework of the KMM and makes use of the approach of the loop
equations.

The main object of this approach is the following generating function
\footnote{Note that the second scalar factor contains $\bar\Phi$.}
\ba
G_{\alpha\beta} = \langle {\rm Tr}\frac{1}{\alpha I - \Phi}
U\frac{1}{\beta I - \bar\Phi}U^\dagger \rangle / \langle 1
\rangle  =
\non
= \int d\gamma\rho(\gamma) d\bar\gamma \bar\rho(\bar\gamma)
\frac{C_{\gamma\bar\gamma}}{(\alpha-\gamma)(\beta-\bar\gamma)}
{}~.
\label{defG}
\ea
It satisfies the following loop equation \cite{DMS}
\be
(E_\alpha + \beta)G_{\alpha\beta} - E_{\alpha} \stackrel{{\rm l.e.}}{=}
\oint\frac{d\omega}{2\pi i}\frac{L(\omega)G_{\omega\beta}}
{\alpha - \omega} \equiv L(\alpha)G_{\alpha\beta} - R_{\beta}(\alpha);
\nn \\ (\bar E_\beta + \alpha)G_{\alpha\beta} - \bar E_{\beta}
\stackrel{{\rm l.e.}}{=}
\oint\frac{d\bar\omega}{2\pi i}\frac{\bar L(\bar\omega)G_{\alpha\bar\omega}}
{\beta - \bar\omega} \equiv \bar L(\beta)G_{\alpha\beta} -
\bar R_{\alpha}(\beta)
\label{loopeq}
\ee
where
\be
E_\alpha = {\langle {\rm Tr}\frac{1}{\alpha I - \Phi} \rangle}
/{\langle 1 \rangle},\ \ \bar E_\beta = {\langle {\rm
Tr}\frac{1}{\beta I - \bar\Phi} \rangle}/ {\langle 1 \rangle}~,
\nonumber
\ee
so that ${\rm Disc}_\alpha E_\alpha = 2\pi i \rho(\beta)$, while ${\rm
Disc}_\beta \bar E_\beta = 2\pi i \bar\rho(\beta)$.  Functions
$L(\omega)$ and $\bar L(\bar\omega)$ are some functions. On the
support of the corresponding spectral density, these two functions are
sums of the derivative of the logarithm of the Itzykson-Zuber integral
and the real part of the generating function $E(\bar E)$. Both $L$ and
$\bar L$ are analytic in the vicinity of the supports of $\rho$ and
$\bar\rho$ and in principle can have cuts and singularities elsewhere.
For the sake of simplicity we assume below that these singularities
are located at $\omega = \infty$ and $\bar\omega = \infty$, and
neither $L$ nor $\bar L$ have cuts. These two functions are a very
natural starting point for the DMS procedure. If one assumes an ansatz
for $L$ and $\bar L$, then the search for the corresponding $E_\alpha$
and $C_{\alpha\beta}$ can be reduced to an algebraic problem
\cite{DMS,Penner}. Here, however, we will offer another variation of
DMS approach, which is closer to the Matytsin's ideas of computing the
Itzykson-Zuber integral and considers instead of $L,\bar L$ their
combination with $E_\alpha,\bar E_\alpha$ \re{g-EL}.

The functions $R_\beta(\alpha)$ and $\bar R_\alpha(\beta)$, enetering
right hand sides of \eq{loopeq}, are defined as the contributions to
the integrals (\ref{loopeq}) from the residues at infinity and
possible singularities of the functions $L$ and $\bar L$.  According
to its definition $R_\beta(\alpha)$ has no discontinuities as a
function of $\alpha$, ${\rm Disc}_\alpha R_\beta(\alpha) = 0$.
Moreover, as a function of $\alpha$ it has its singularities only at
those of $L(\alpha)$ (which are actualy singularities of $g_R(\alpha)$
- see eq.(\ref{g-EL})).  As a function of $\beta$, $R_\beta(\alpha)$
is analytic outside the support of $\brho$, where its imaginary part
has a finite jump. Under our simplifying assumption this means that
$R_\beta(\alpha)$ is a polynomial in positive powers of $\alpha$ with
$\beta$-dependent, everywhere finite, coefficients.\footnote{If
$L(\alpha)$ has extra singularities at some point $\nu$,
$R_\beta(\alpha)$ will also contain a contribution, which is
polinomial in $(\alpha - \nu)^{-1}$. This correction is easy to be
accounted for in our reasoning below.}

{}From (\ref{loopeq}) it follows that
\be
G_{\alpha\beta} = \frac{E_\alpha - R_{\beta}(\alpha)} {\beta -
g_R(\alpha)} = \frac{\bar E_\beta - \bar R_{\alpha}(\beta)} {\alpha -
g_L(\beta)}~,
\label{G-g}
\ee
where
\be
g_R(\alpha) = -E_\alpha + L(\alpha)~, \nn \\ g_L(\beta) = -\bar
E_\beta + \bar L(\beta)~.
\label{g-EL}
\ee
According to their definitions $E_\alpha \sim \frac{1}{\alpha}$, while
$G_{\alpha\beta} \sim \frac{\bar E_\beta}{\alpha}$ as $\alpha
\rightarrow
\infty$.

Then $C_{\alpha\beta}$ is nothing but a double discontinuity of
$G_{\alpha\beta}$:
\be
C_{\alpha\beta} = \frac{{\rm Disc}_\alpha}{2\pi i \rho(\alpha)}
\frac{ {\rm Disc}_\beta}{2\pi i \bar\rho(\beta)} G_{\alpha\beta} =
 -\frac{{\rm Disc}_\beta R_\beta(\alpha) /2\pi i
\bar\rho(\beta)}{(\beta - g_R(\alpha)) (\beta - g_R^*(\alpha))}~,
\label{C-R}
\ee
where we have defined on the cut
\be
g_R(\alpha) = L(\alpha)-{1\over 2} \tilde V^\prime (\alpha)
+i\pi\rho(\alpha)~, ~~{1\over 2} \tilde V^\prime (\alpha) \equiv {\rm
Re} E_\alpha~,
\label{defg}
\ee
and ``*'' denotes complex conjugate.  Alternative expression is
\be
C_{\alpha\beta} = -\frac{{\rm Disc}_\alpha \bar R_\alpha(\beta) /2\pi
i \rho(\alpha)}{(\alpha - g_L(\beta)) (\alpha - g_L^*(\beta))}~.
\label{C-L}
\ee
Assume that the correlator $C_{\alpha\beta}$ can be naively
analytically continued (i.e. just by using instead of $\alpha$ and
$\beta$ two arbitrary complex numbers) from the cut.  Since for any
given $\alpha$ the positon of the poles of the expressions
\re{C-R} and \re{C-L} should concide we find that
\be
g_L (g_R^*(\alpha))=\alpha~,~~g_R (g_L^*(\beta))=\beta~.
\label{inverse}
\ee
Thus we recover the functional form of the Master Field Equation in
the Matytsin's approach \cite{Ma}.

It remains to define the quantity in the numerator in (\ref{C-R}).  In
order to do this let us return to eq.(\ref{G-g}) and rewrite it as
\footnote{
We use the fact, following from the definition \re{defG}, that
\be
\frac{1}{2\pi i}{\rm Disc}_\beta G_{\alpha\beta} =
\int \frac{\rho(\gamma)d\gamma}{\alpha - \gamma}
C_{\gamma\beta}\bar\rho(\beta).
\nn
\ee
}
\be
\frac{{\rm Disc}_\beta R_\beta(\alpha)}{2\pi i \bar\rho(\beta)} =
(\beta - g_R(\alpha))
\frac{{\rm Disc}_\beta G_{\alpha\beta}}{2\pi i \bar\rho(\beta)}
= \int \rho(\gamma)d\gamma C_{\gamma\beta}
\left[\frac{\beta - g_R(\alpha)}{\alpha-\gamma}\right].
\ee

Now it is time to recall that $R_\beta(\alpha)$ and thus ${\rm
Disc}_\beta R_\beta(\alpha)$ is a polynomial in positive powers of
$\alpha$ (provided $g_R$ and $L$ are singular only at infinity).  This
allows to substitute $\left[\frac{\beta -
g_R(\alpha)}{\alpha-\gamma}\right]$ at the r.h.s. by the part of its
asymptotics in $\alpha$, containing only its positive powers
\be
\left.\left[\frac{\beta - g_R(\alpha)}{\alpha-\gamma}\right]\right._+
= - \left.\left[\frac{g_R(\alpha)}{\alpha-\gamma}\right]\right._+
\equiv -\sum_{k,l\geq 0} \alpha^k\gamma^l \sigma_{kl}
\label{sigmas}
\ee
i.e. by a finite polinomial in $\alpha$ and $\gamma$ with coefficients
$\sigma_{kl} = \sigma_{kl}\{g_R\}$, totally defined by the shape of
the function $g_R(\alpha)$.

Thus
\be
\frac{{\rm Disc}_\beta R_\beta(\alpha)}{2\pi i \bar\rho(\beta)} =
- \sum_{k,l\geq 0} \alpha^k \sigma_{kl} M_l(\beta),
\ee
where
\be
M_l(\beta) = \int \gamma^l \rho(\gamma) d\gamma C_{\gamma\beta}.
\label{moments}
\ee
Note that it follows from the normalization condition for
$C_{\alpha\beta}$ that \be M_0(\beta)=1
\label{M0}
\ee
Finally, from (\ref{C-R}) we obtain:
\be
C_{\alpha\beta} = -\frac{\sum_{k,l\geq 0} \alpha^k M_l(\beta)
\sigma_{kl}\{g_R\}} {(\beta - g_R(\alpha)) (\beta - g_R^*(\alpha))},
\label{C-M}
\ee
and $M_l(\beta)$ in this formula can be obtained from solution of a
finite system of equations, which arises after (\ref{C-M}) is
resubstituted into (\ref{moments}) with the condition \re{M0}.

This provides a complete solution for the problem of evaluation of
$C_{\alpha\beta}$, provided one starts from any adequate (i.e.
satisfying (\ref{inverse})) pairs of functions $g_{L,R}(\alpha)$.  We
did not {\it prove} here that {\it any such} pair of functions
$g_{L,R}(\alpha)$ provides an answer: we rather proved that any answer
has this form with some $g_{L,R}(\alpha)$. It is clear, however, that
further restrictions on the shape of $g_R$ should {\it not} be
imposed: otherwise it would be impossible to suite arbitrary densities
$\rho$ and $\bar\rho$. Therefore the alrorithm is obliged to work.

Unfortunately, in all but the simplest cases the analytic structure of
$L(\alpha)$ and/or $E_\alpha$ is very complicated, which hinders any
practical calculations. One can see it by following. Take {\it any}
function $g_L^*$, which has negative imaginary part when its argument
belongs to some real interval (support of $\rho$). Then consider the
inverse of this function and call it $g_R$ (it must have an interval
on the real axis, the support of $\rho$, where its imaginary part is
positive). In most cases the analytic structure of $g_R$ turns out
very involved.

For illustrative purposes we turn to the bi-semi-circle example, where both
$\rho$ and $\brho$ are semi-circle.
We start from the pair of functions
\ba
g_L(\alpha) = \frac{\sqrt{\mu^2+4\mu/\bmu}}{2}\alpha
+\sqrt{\frac{\mu^2\alpha^2}{4}-\mu}
\non
g_R (\beta) =
\frac{\sqrt{\bmu^2+4\bmu/\mu}}{2}\alpha+\sqrt{\frac{\bmu^2\alpha^2}{4}-\bmu}
\ea
First let us find the coefficients $\sigma_{kl}$ entering the power
expansion \re{sigmas}:
\be
\sigma_{00} =-{1\over 2} (\sqrt{\mu^2+4\mu/\bmu}+\mu)~,~~\sigma_{k>0,l>0}=0
\ee
The general result for the correlator \re{C-M} then takes the form
\be
C_{\alpha\beta}(1\vert \rho_\mu, \rho_{\bar\mu}) = {1\over 2}
\frac{\sqrt{\mu^2\bmu^2+4\mu\bmu}+\mu\bmu}{\mu \alpha^2+\bmu\beta^2 -
\sqrt{\mu^2\bmu^2+4\mu\bmu}\alpha\beta +\mu\bmu~},
\label{C-M-gauss}
\ee
which coincides with the result of \cite{DMS}.

%\centerline{\it Evaluation of $C(0\vert \rho,\bar\rho)$}

Knowledge of the correlator allows one in principle to compute the
Itzykson-Zuber integral itself. To demonstrate that let us introduce
an auxilary dependence of $C(n)$ on some parameter $t$ in the
following fashion
\be
C_t(0) = {1\over N^2} \log \int [dU] e^{Nt \tr \Phi U \bar\Phi U^\dagger}~.
\ee
Then for its derivative we find
\ba
& t \frac{\partial C_t(0)}{\partial t} =
\frac{\int [dU] {t\over N} \tr\Phi U \bar\Phi U^\dagger
e^{Nt \tr \Phi U \bar\Phi U^\dagger} }{ \la 1 \ra}=
\non
&= \int d\alpha \rho_t(\alpha) \alpha d\beta \bar\rho_t(\beta)
\beta C_{\alpha\beta}(t)
\ea
As an example let us consider again the case of the bi-semi-circle
distribution.  The dependence on $t$ can be absorbed into the
redefinition $\mu$ and/or $\bar\mu$. It is convenient to consider the
symmetric redefinition
\be
\mu \rightarrow{\mu\over \sqrt t}~,~~\bar\mu
\rightarrow{\bar\mu\over \sqrt t}~.
\ee
Then
\be
 t \frac{\partial C_t(0)}{\partial t} =
\frac{2t^2}{\sqrt{\mu\bar\mu}(\sqrt{\mu\bar\mu +4t^2}+\sqrt{\mu\bar\mu})}
\ee
Integrating the last line one obtains the generalization
\re{Gross2} of the Gross' result
\be
C(0\vert \rho_\mu, \rho_{\bar\mu}) =
\frac{\sqrt{\mu\bar\mu}-\sqrt{\mu\bar\mu}}{2\sqrt{\mu\bar\mu}} -
{1\over 2}
\log\frac{\sqrt{\mu\bar\mu}+\sqrt{\mu\bar\mu}}{2\sqrt{\mu\bar\mu}}
\label{Gross2}
\ee

\bigskip

\centerline{{\it  Matytsin's theory} }

\bigskip

A more direct way of calculating the Itzykson-Zuber integral $C(0)$,
which does not require the knowledge of the correlators $C(1)$, was
developed recently in \cite{Ma}.  It has been proved that
\ba
&C(0\vert \rho,\bar\rho) = S(\rho,\bar\rho) +
\frac{1}{2}\int \rho(\phi)\phi^2 d\phi +
\frac{1}{2}\int \bar\rho(\phi)\phi^2 d\phi -
\nn \\
&-\frac{1}{2}\int\int \rho(\phi)\rho(\phi')\log(\phi-\phi')d\phi
d\phi' -\frac{1}{2}\int\int \bar\rho(\phi)\bar\rho(\phi')
\log(\phi-\phi')d\phi d\phi'.
\label{Ma-Itzykson-Zuber}
\ee
Integrals are over supports of $\rho(\phi)$ and $\bar\rho(\phi)$ and
\be
S(\rho,\bar\rho) = \frac{1}{2}\int_0^1 dt \int d\phi
[{\rm Im}f(\phi,t)]\left([{\rm Re}f(\phi,t)]^2 + \frac{1}{3}
[{\rm Im}f(\phi,t)]^2\right).
\label{S-Ma}
\ee
The function $f(\phi,t)$ is a solution to the ``Hopf
equation''\footnote{ It is a remarkable equation, the first one in the
``quasiclassical KdV hierarchy'', which plays an important in the
theory of topological Landau-Ginsburg models and Generalized
Kontsevich model. In Itzykson-Zuber theory $1/t$ plays the role of the
coefficient in front of the action, $\frac{1}{t}{\rm Tr}\Phi U\bar\Phi
U^\dagger$.  It is interesting to understand the role of other KdV
times in this context.  The ``action'' $S(\rho,\bar\rho)$ in
(\ref{S-Ma}) is {\it not} an action for Hopf equation - it is rather
one of the (quasiclassical-)KdV Hamiltonians. We note in passing that
the action for the Hopf equation is not unique: see ref.\cite{FGM}
(where it is refered to as Bateman equation) for discussion of the
corresponding universality structure.  }
\be
\frac{\partial f}{\partial t} = \phi\frac{\partial f}{\partial \phi}
\label{Hopf}
\ee
with the initial condition
\be
f(\phi,t=0) = g_R(\phi) - \phi.
\label{inicond}
\ee
It is then a consequence of the Hopf evolution that
\be
f(\phi,t=1) = -g_L(\phi) + \phi.
\label{anobc}
\ee
The function $f(\phi,t)$ is in fact determined from a {\it algebraic}
equation, because the differential Hopf equation \re{Hopf} is in fact
explicitly integrable. Its generic solution is given in the parametric
form:
\be
\phi = \alpha + tF(\alpha), \nn \\
f(\phi,t) = F(\alpha),
\ee
with arbitrary function $F(\alpha)$. The shape of $F(\alpha)$ is
dictated by initial condition (\ref{inicond}):
\be
F(\phi) = f(\phi,t=0) = g_R(\phi) - \phi.
\ee
Given this function, one finds $\alpha(\phi,t)$ from the algebraic
equation
\be
\phi = \alpha(\phi,t) + tF\left(\alpha(\phi,t)\right),
\label{eqforalp}
\ee
and then obtains
\be
f(\phi,t) = F\left(\alpha(\phi,t)\right) =
g_R\left(\alpha(\phi,t)\right) - \alpha(\phi,t).
\label{f(phi,t)}
\ee
In order to make this (exhaustive) description a little bit more
transparent we turn to the semi-circle case.
\ba
&\rho(\alpha) = \bar\rho(\alpha) = \rho_\mu(\alpha) =
\frac{1}{\pi}\sqrt{\mu - \frac{\mu^2\alpha^2}{4}}~;
\nn \\
& g_R(\alpha) = \alpha\frac{\sqrt{\mu^2 + 4}}{2} +
i\pi\rho_\mu(\alpha) = \sqrt\mu\left(\hat\alpha
\frac{\cosh\tau}{\sinh\tau} + i \sqrt{1-\hat\alpha^2}\right)~,
\nn \\
& g_L(\beta) = \beta\frac{\sqrt{\mu^2 + 4}}{2} + i\pi\rho_\mu(\beta) =
\sqrt\mu\left(\hat\beta \frac{\cosh\tau}{\sinh\tau} - i
\sqrt{1-\hat\beta^2}\right)~,
\nn \\
& \alpha = \frac{2}{\sqrt\mu}\hat\alpha = \sqrt\mu\frac{\hat\alpha}{\sinh\tau},
\ \ \beta = \sqrt\mu\frac{\hat\beta}{\sinh\tau}~.
\ea
Then
\be
F(\phi) = f(\phi,t =0) = g_R(\phi) - \phi = \sqrt\mu\left(
\frac{\cosh\tau -1}{\sinh\tau}\hat\phi + i\sqrt{1-\hat\phi^2}\right).
\ee
Given such $F$ (\ref{eqforalp}) is actually a quadratic equation for
$\alpha(\phi,t)$, which is easily solved:
\be
\alpha(\phi,t) =
\frac{\phi (1+t(\cosh\tau-1)) +i\sqrt{1+2t(1-t)(\cosh\tau-1)-\phi^2}}
{1+2t(1-t)(\cosh\tau-1)}
\ee
and substituting this into (\ref{f(phi,t)}) we get:
\be
f(\phi,t) = \sqrt\mu
\frac{\frac{\cosh\tau -1}{\sinh\tau}(1-2t)\hat\phi +
 i\sqrt{1+2t(1-t)(\cosh\tau - 1) - \hat\phi^2}} {1+2t(1-t)(\cosh\tau -
1)}.
\ee
(One can now see that $f(\phi,t=1) = \sqrt\mu\left( -\frac{\cosh\tau
-1}{\sinh\tau}\hat\phi + i\sqrt{1-\hat\phi^2}\right)$ is indeed equal
to $-g_L(\phi)+\phi$, in accordance with (\ref{anobc}).)

Integral (\ref{S-Ma}) for $S(\rho_\mu,\rho_\mu)$ is now easy to
evaluate and we get:
\be
S(\rho_\mu,\rho_\mu) = -\frac{\sqrt{\mu^2+4}}{2\mu} + \frac{1}{\mu} +
\frac{1}{2}
\log\frac{\sqrt{\mu^2+4}+\mu}{2}.
\ee

To find the Van der Monde determinant in the case of the
bi-semi-circle distribution, we first note that it is easy to
calculate the following integral
\be
\int\int \rho_{s,\mu_1}(\alpha)\rho_{s,\mu_2}(\beta) \log(z\alpha-\beta)
d\alpha d\beta = \nn \\ =\left\{
\begin{array}{ll}
z^2 \leq \frac{\mu_1}{\mu_2}, \ \ & \frac{z^2}{4}\frac{\mu_2}{\mu_1} -
\frac{1}{2}(1 + \log\mu_2) \\ z^2 \geq \frac{\mu_1}{\mu_2}, \ \ &
\frac{1}{4z^2}\frac{\mu_1}{\mu_2} -
\frac{1}{2}\left(1 + \log\frac{\mu_1}{z^2}\right) \end{array}
\right.
\ee
Thus, the logarithm of the Van der Monde determinant for the
semi-circle distribution is
\be
\int\int \rho_\mu(\phi)\rho_\mu(\phi')\log(\phi - \phi')d\phi
d\phi' = -\frac{1}{2}\log\mu -\frac{1}{4}
\ee

Recalling that
\be
\int\rho_\mu (\phi)\phi^2d\phi  = \frac{1}{\mu}~, ~~
\int\brho_{\bmu} (\phi)\phi^2d\phi  = \frac{1}{\bmu}~,
\ee
we reproduce Gross's answer \cite{Gross} for semi-circle distribution,
\be
C_\mu = \frac{\sqrt{\mu^2+4}-\mu}{2\mu} -
\frac{1}{2}\log\frac{\sqrt{\mu^2+4}+\mu}{2\mu}.
\ee

We aslo note in passing that the item $S(\rho,\bar\rho)$ at the r.h.s.
of (\ref{Ma-Itzykson-Zuber}) can be distinguished from the other terms
for the following reason.  Exact answer for the Itzykson-Zuber
integral is:
\be
\frac{1}{V_N}\int [dU] e^{{\rm Tr}\Phi U \bar\Phi U^\dagger} =
\frac{{\rm det}_{ij} e^{\phi_i\bar\phi_j}}{\Delta(\phi)\Delta(\bar\phi)} =
\nn \\
= \frac{e^{\frac{1}{2}{\rm Tr}\Phi^2 + \frac{1}{2}{\rm Tr}\bar\Phi^2}}
{\Delta(\phi)\Delta(\bar\phi)} {\rm det}_{ij} e^{-\frac{1}{2}(\phi_i-
\bar\phi_j)^2}
\ee
It is the last determinant at the r.h.s. which corresponds to
$e^{S(\rho,\bar\rho)}$ in the large $N$ limit. This determinant tends
to unity for very broad distributions when the average distance
between adjacent eigenvalues is large. In the example of semi-circle
distributions this corresponds to $\mu,\bar\mu
\longrightarrow 0$ and indeed
$S(\rho,\bar\rho) \sim O(1)$, while the other terms are
$\sim O(1/\mu,\log\mu)$.


\begin{thebibliography}{12}

\bibitem{KM} V.Kazakov and A.Migdal, {\it Nucl.Phys.} {\bf B397} (1993) 214
\bibitem{itzub} Harish-Chandra, {\it Amer. J. Math.} {\bf 79} (1957) 87;\\
C.Itzykson and J.B.Zuber, {\it J. Math. Phys.} {\bf 21} (1980)  411
\bibitem{Migdal} A.Migdal, {\it Mod.Phys.Lett.} {\bf A8} (1993) 359
\bibitem{DMS} M.Dobroliubov, Yu.Makeenko and G.Semenoff, {\it Mod.Phys.Lett.}
{\bf A8} (1993) 2387
\bibitem{Gross} D.Gross, {\it Phys.Lett.} {\bf 293B} (1992) 181
\bibitem{Penner} Yu.Makeenko, {\it Phys.Lett.} {\bf B314} (1993) 197
\bibitem{ZN} I.Kogan, G.Semenoff and N.Weiss, {\it Phys. Rrev. Lett.} {\bf 69}
(1992) 3435
\bibitem{KMSW} I.Kogan et al. {\it Nucl. Phys.} {\bf B395} (1993) 547
\bibitem{Migdal-mixed} A.Migdal, {\it Mod. Phys. Lett.} {\bf A8} (1993) 245
\bibitem{DKSW} M.Dobroliubov et al. {\it Phys.Lett.} {\bf 302B} (1993) 283
\bibitem{morev}A. Morozov,  {\it Integrability and Matrix Models}, preprint
ITEP-M2/93, ITFA 93-10, 1993
\bibitem{boulkaz}D.Boulatov and V.Kazakov, {\it Int. J. Mod. Phys.} {\bf A8}
(1993) 809
\bibitem{MoSha}S.Shatashvili, {\it Comm. Math. Phys.} {\bf 154} (1993) 421; \\
 A.Morozov, {\it Mod. Phys. Lett.} {\bf A7} (1992) 3503
\bibitem{Ma} A.Matytsin, {\it On the Large N Limit of the Itzykson-Zuber
Integrall}, preprint  PUPT-1405/93; hep-th/9306077
\bibitem{brook} S. Aoki and A. Gocksch, {\it Nucl. Phys.} {\bf B404} (1993) 173
\bibitem{FGM} D.Fairlie, J.Govaerts et al., {\it Nucl.Phys.} {\bf B373} (1992)
214

\end{thebibliography}
\end{document}